\documentclass[prl,nofootinbib,preprintnumbers]{revtex4}

\usepackage{amsfonts}
\usepackage{mathrsfs}
\usepackage{amssymb}
\usepackage{amsmath}
\usepackage{graphicx,epsfig}

\def\lsim{\mathrel{\mathpalette\@versim<}}

\def\gsim{\mathrel{\mathpalette\@versim>}}

\begin{document}

\preprint{CERN-PH-TH/2010-087, ACT-05-10, MIFP-10-16}

\title{Comments on Ultra-High-Energy Photons and D-Foam models}

\author{John Ellis}
\affiliation{CERN, Theory Division, CH 1211 Geneva 23, Switzerland.}

\author{Nick E. Mavromatos}
\affiliation{King's College London, Department of Physics, London WC2R 2LS, U.K.}

\author{Dimitri V. Nanopoulos}

\affiliation{George P. and Cynthia W. Mitchell Institute for
Fundamental Physics,
Texas A$\&$M University, College Station, TX 77843, USA }

\affiliation{Astroparticle Physics Group,
Houston Advanced Research Center (HARC),
Mitchell Campus, Woodlands, TX 77381, USA}

\affiliation{Academy of Athens, Division of Natural Sciences,
28 Panepistimiou Avenue, Athens 10679, Greece }

\begin{abstract}

We revisit the constraints that the non-observation of ultra-high-energy photons due to the
GZK cutoff can impose on models of Lorentz violation in photon propagation, following
recent work by Maccione, Liberati and Sigl [arXiv:1003.5468] that carries further an earlier
analysis by the present authors (Phys. Rev. D {\bf 63},12402 (2001), [hep-th/0012216]). We argue
that the GZK cutoff constraint is naturally evaded in the D-brane model of space-time foam
presented recently by the present authors (Phys.\ Lett.\  B {\bf 665}, 412 (2008) [arXiv:0804.3566]), in
which Lorentz-violating effects on photon propagation are independent of possible effects
during interactions. We also note a novel absorption mechanism that could provide a
GZK-like cutoff for photons in low-scale string models.

\end{abstract}

\maketitle


There has been much discussion~\cite{UHELIV} of the possible implications of Lorentz violation
for ultra-high-energy cosmic rays and the Greisen-Zatsepin-Kuzmin (GZK) cutoff~\cite{GZK}. In
particular, it was observed that the GZK cutoff might be removed for ultra-high-energy
photons striking the cosmic microwave background~\cite{Kifune} and/or the astrophysical infra-red
background~\cite{PM}. The present authors discussed this issue~\cite{EMNGZK} in the context of a space-time
foam model based on recoiling D-branes~\cite{Dfoam2}, proposing a formalism that has
recently been used in an analysis by Maccione, Liberati and Sigl~\cite{libsigl}. These authors
argue that the apparent observation of a GZK cutoff for cosmic-ray primaries~\cite{Auger},
combined with the non-observation of ultra-high-energy photons~\cite{Augergamma}, is strong
circumstantial evidence that the GZK mechanism is also at work for photons, and show that this
imposes very strong constraints on the class of Lorentz-violating models considered in~\cite{EMNGZK}.

Since the publication of~\cite{EMNGZK}, however, we have developed a new
class of D-particle models for
space-time foam~\cite{Dfoam,Dfoam2,emnnewuncert,li}, within which the GZK constraint requires
re-examination. In these models, space-time is punctuated by defects that may be either
(i) point-like D0-branes (D-particles)~\cite{Dfoam}, with no electric charge, in which case only electrically neutral matter and radiation (represented as open strings with their ends attached on the brane world representing our Universe) can interact non-trivially with the foam, or (ii) D3-branes
wrapped up around small three-cycles so as to resemble small
spheres (`effective D-particles'), permitting charge flow on their surfaces. In the latter case, there could be interactions of charged matter with the D-defects, which however are suppressed compared to photons, for the purely stringy reasons discussed in~\cite{li}. Hence, the only high-energy processes where Lorentz
violation might be relevant are those involving photons, \emph{e.g.}, $\gamma + \gamma_{\rm B}
\rightarrow e^+ + e^-$ where $\gamma_{\rm B}$ is a cosmic microwave or infrared background photon.
The hadronic GZK cutoff processes, \emph{e.g.}, $p + \gamma_{\rm B} \rightarrow \Delta ^+ \rightarrow p + \pi^0 \quad {\rm or} \quad n + \pi^+$, involve an incident charged high-energy particle ($p$),
which has vanishing (or suppressed) interactions with the D-particles, in which case
the GZK cutoff is unaffected, as indicated by experiment~\cite{Auger,Augergamma}.

In these D-particle models, the propagation of photons is affected as originally
suggested in~\cite{AEMNS}, with the arrival times of photons delayed by
amounts proportional to their energies $E$, due to an energy-dependent average phase shift:
$exp(i (E + c'.E^2).t - {\bf p}.{\mathbf x}))$, where $c'$ is a proportionality factor 
parametrizing Lorentz violation, that depends in general on the density of foam particles as well as
the microphysical model~\cite{emnnewuncert,li}. On the other hand,
particle interactions conserve
Lorentz-invariantly both energy $E$ and momentum ${\bf p}$, in a leading approximation. Thus,
there is no direct connection between Lorentz violation in the {\it propagation} of photons and in
their {\it interactions}. As we show below, it is not possible, in general, to constrain possible time-delay
parameters by considerations of the GZK cutoff: the very interesting constraints they impose
on Lorentz violation~\cite{libsigl} are evaded by the D-particle models
of~\cite{Dfoam,Dfoam2,emnnewuncert,li}. We shall make these points clearer
in the following discussion, where we consider first D-foam effects on photon
propagation and subsequently those on interactions, \emph{i.e}., we first consider possible effects
on 2-point Green functions and subsequently 1-particle irreducible (1PI) higher-point
functions.

The effect of D-particle foam on photon propagation {\it in vacuo}
has analogies with the appearance of a refractive
index for photons propagating through a material medium~\cite{emnnewuncert}.
In that case, photons interact with atomic electrons of effective mass $m$,
which may be modelled~\cite{feynman}
as simple-harmonic oscillators with a resonant frequency $\omega_0$. In the presence
of a photon of frequency $\omega$, these are subject to an oscillating external electric force:
$F= e E_0 e^{i \omega t}$, where $e$ is the electron charge. The electrons are excited by the
equation of motion $m \left(d^2x/dt^2 + \omega_0^2 x \right) = e E_0 e^{i \omega t}$,
inducing an electric field
$ E_a = - (e n_e/\epsilon_0) i (e E_0/[m (\omega^2 - \omega_0^2)])e^{i\omega (t - z)}~,$
where we use units in which the unmodified velocity of light $c=1$, 
$\epsilon_0$ is the dielectric constant of the 
vacuum and $n_e$ is the area density of
electrons in the medium. As a consequence, light propagates through the medium at a
speed $1/n$, where $n$ is the refractive index, causing a delay $\Delta t$
while traversing a distance $\Delta z$ given by
$\Delta t = (n - 1) \Delta z$. Representing the electric field before and
after passing through a medium of thickness $\Delta z$ as
$E_{\rm before} = E_0 e^{i\omega (t - z)}$ and
$E_{\rm after} = E_0 e^{i\omega (t - z - (n-1) \Delta z)}$, in the case
of small deviations from the vacuum refractive index
we have $E_{\rm after} \simeq E_0 e^{i \omega(t - z)} - i[\omega(n -1)\Delta z]
E_0 e^{i\omega(t-z)}$, where the last term on the right-hand-side of this relation
is just the field $E_a$ produced by the oscillating electrons after passing through the medium.
Thus we find the standard formula for the refractive index in a conventional medium:
\begin{equation}
  n = 1 + \frac{{\rho_e} e^2 }{2\epsilon_0 m (\omega_0^2 - \omega^2)} ,
  \label{refrordinary}
  \end{equation}
where $\rho_e$ is the density of electrons.
We see in (\ref{refrordinary}) that the refractive index in an ordinary medium is inversely proportional to (the square of) the frequency $\omega$ of light, as long as it smaller than the oscillator
frequency, where the refractive index diverges. Notice that physical recoil of the electrons
during scattering with photons is not relevant in the derivation of the refractive index (\ref{refrordinary}).

In the D-foam models~\cite{Dfoam,Dfoam2,emnnewuncert,li}, 
the r\^ole of the electrons of the material medium
is played by the D-particles themselves.
In our preferred formulation of the D-foam, as discussed in~\cite{emnnewuncert}, 
when a photon strikes a D-particle
it creates an intermediate string between the D-particle and the D3-brane
on which the photon moves. This string stores the energy $E$ of the photon as potential
energy, by stretching to a length $L$ and acquiring $N$ internal oscillator
excitations, cf, the electron oscillators discussed above:
\begin{equation}
E \; = \; \frac{L}{\alpha'} + \frac{N}{L} .
\label{string1}
\end{equation}
The maximal string length is $L_{\rm max} = \alpha' E/2$,  and the time
taken by the string to grow to this length and then shrink back to its minimal
size is
\begin{equation}
\Delta t \; \sim \; \alpha' E.
\label{string2}
\end{equation}
This is then the order of magnitude of the time delay
experienced by a photon interacting with a single D-particle. Using the same relation
$\Delta t = (n - 1)\Delta z/c$ as in the case of a conventional medium, we infer that the
refractive index $n$ for a photon propagating through a gas of D-particles of density $\rho_D$
is of order
\begin{equation}
n (E) \; = \; 1 + \alpha' E \sigma \rho_D
\label{string3}
\end{equation}
where $\sigma$ is the photon-D-particle scattering cross section.
In this case, we see in (\ref{string3}) that the refractive index in a D-particle medium is
proportional to the energy $E$ of the photon, at least as long as it smaller than the string
energy scale. At energies comparable to the D-particle mass scale, 
a more complete estimate is required, and there may be an analogue of the
divergence in (\ref{refrordinary}), as we discuss below.
Notice also, that as in the derivation of (\ref{refrordinary}), physical recoil of the D-particle
during scattering with photons is not directly relevant in the derivation of the refractive index
(\ref{string3}).

We had previously given a heuristic argument for the formula (\ref{string3}),
based on the observation that collisions of photons with D-particles would in general
cause the latter to recoil, modifying the effective metric
`felt' by the photons during propagation and giving it the off-diagonal form:
\begin{equation}
G_{\mu\nu} = \eta_{\mu\nu} + h_{\mu\nu} ~, \qquad
h_{0i} = u_i \ll 1~,
\label{Finsler}
\end{equation}
where
\begin{equation}\label{recvel}
u_i = \frac{g_s}{M_s} \Delta p_i,
\end{equation}
is the velocity of the recoiling D-particle following scattering by the photon through a
momentum transfer $\Delta p_i \equiv r p_i$. 
Here $g_s$ the string coupling, $M_s$ is the string scale, and
the D-particle mass $\sim M_s/g_s$. It was assumed in~\cite{EMNGZK} and in our early models of
D-foam~\cite{Dfoam2} that
there is an average background recoil velocity field $\ll u_i \gg = g_s\frac{r}{M_s} p_i \ne 0 $ over the collection of D-particles encountered by the photon. The expression (\ref{Finsler}) is a
Finsler metric, in the sense that it depends on the momentum $p_i$ of the particle,
which leads (on average) to a modified dispersion relation for a massless particle such as the
photon:
\begin{equation}
p^\mu p^\nu G^{\mu\mu} = E^2 + {\mathbf p}^2 + 2\xi \, E\, {\mathbf p}^2 = 0~: \qquad \xi \equiv g_s \frac{r}{M_s} .
\label{modified}
\end{equation}
Such a modified dispersion relation would lead to a non-trivial refractive index, $n - 1$ of the form
\begin{equation}
n \; = \; 1 + \mathcal{O}\left(\xi\,|{\mathbf p}|\right) \, > \, 1~,
\end{equation}
where $\xi \sim \ll u_i \gg$.
It is the quantity $\xi$ that was argued in~\cite{libsigl} to to be constrained by $\xi < 10^{-12}$.
However, this constraint is evaded in isotropic foam models with $\ll u_i \gg = 0$, and in particular it
does not apply to the stretched-string model discussed above, for the reasons we discuss
below.

To understand the impact of the constraint due to~\cite{libsigl}, we first recall that it
was based on an analysis of 1PI (amputated) scattering amplitudes involving photons.
In conventional quantum field theory, energy-momentum is conserved in any scattering
process, but this assumption needs to be re-examined within any Lorentz-violating
framework, as was pointed out in~\cite{EMNGZK}. At our present
level of ignorance, for the phenomenological purposes of establishing constraints~\cite{libsigl},
Lorentz violation and/or energy non-conservation in a 1PI scattering
amplitude should be considered independently from Lorentz violation during particle
propagation. 

In general in our approach, since the tree-level 1PI amplitude for
$\gamma + \gamma_{\rm B} \rightarrow e^+ e^-$ involves an internal electron, not a photon,
and Lorentz violation is absent for electrons in our D-foam model, it is absent in the 
tree-level amplitude. Possible effects due to virtual photons in loop corrections need to to be
investigated, but it is clear that Lorentz violation must be strongly suppressed, if it appears at all.
Moreover, within the framework (\ref{string1}, \ref{string2}, \ref{string3})
of string formation, stretching and decay that we now prefer~\cite{emnnewuncert},
there is also no mechanism for momentum non-conservation in 1PI amplitudes.
Furthermore, the appropriate kinematics is that the external legs  should all
be regarded as on mass-shell. This is to be contrasted with the framework based
on metric distortion, where momentum is still conserved in 1PI amplitudes, but
where the metric (\ref{Finsler}) should be used for the external photons, with $P_\mu = G_{\mu\nu}P^\nu$
and the modified energy-momentum relation
(\ref{modified}).

The argument of~\cite{libsigl} therefore does {\it not} constrain the string formation
framework (\ref{string1}, \ref{string2}, \ref{string3}) of~\cite{emnnewuncert}, though
it {\it does} impose a severe constraint on the metric distortion framework. However,
we recall that the modification (\ref{modified}) of the standard energy-momentum relation
is proportional to $\ll u_i \gg$, and so could be evaded even in this framework if
$\ll u_i \gg = 0$ (`isotropic foam'). However, even in this case one would expect non-trivial stochastic fluctuations:
\begin{equation}
\label{homog}
\ll u_i u_j \gg \; = \; \sigma^2 \delta_{ij} ~, \qquad \sigma^2 \equiv g_s^2 \frac{\ll r^2 \gg}{M_s^2}~.
\end{equation}
In this case, the recoiling D-particles in the foam provide a background `electric-type' field,
$\vec{u}$, for the $\sigma$-model that describes the open-string excitations corresponding to
photon fields in the first-quantized formalism we use~\cite{mavro2010}. It is
known~\cite{sussk1}  that, in the presence of such electric field backgrounds, there is
space-time non-commutativity, with the string coupling replaced by an effective coupling
\begin{equation}\label{effstringcoupl}
g_s^{\rm eff} = g_s (1 - |\vec u|^2)^{1/2} ~,
\end{equation}
and the space-time metric seen by the photon becomes:
\begin{equation}
\label{openstring}
G_{\mu\nu} = \eta_{\mu\nu} \left(1 - |\vec u|^2 \right)~, \quad \mu, \nu =0, 1~, \qquad
G_{\mu\nu} = \eta_{\mu\nu}~\qquad \mu, \nu \ne 0,\, 1~,
\end{equation}
where we assume that the direction of the recoil is along the $x^1$ coordinate,
and that the space-time is initially flat.
The Finslerian (momentum-dependent) induced metric (\ref{openstring})
depends on the square of the recoil velocity and hence is non-zero even
in isotropic recoil models in which $\ll u_i \gg = 0$.
The average momentum-energy relation of a photon in such a
D-foam background takes the form
\begin{equation}\label{disprel}
p^\mu p^\nu G_{\mu\nu} = 0 = E^2 - {\mathbf p}^2 + \mathcal{O}\left(g_s^2 \frac{E^2 {\mathbf p}^2}{M_s^2} \right)~,
\end{equation}
where the modification is \emph{suppressed quadratically} by the string mass scale,
as a result of (\ref{recvel}).

Notice, however, that both the effective string coupling (\ref{effstringcoupl}) and
the metric (\ref{openstring}) exhibit singular behaviours as $|\vec u|^2 \to 1$,
reminiscent of the singularity in the case (\ref{refrordinary}) of a conventional material medium.
For this reason, caution should be exercised in testing this formalism using
ultra-high-energy cosmic rays in low-scale string models~\cite{pioline}, 
in which it is possible that the photon energy
$E \sim M_s$. The precise forms of the effective string
coupling (\ref{effstringcoupl}) and the metric (\ref{openstring}) are no longer
valid when $|\vec u|^2 = {\cal O}(1)$, as may happen when $E \sim M_s/g_s$.
In fact, for energies above this value, the effective low-energy local field theory description 
\emph{breaks down}~\cite{burgess}, implying that such ultra-high-energy photons
would be \emph{destabilized} when interacting with a D-particle. There would be
a strong space-time distortion leading to \emph{absorption} of such photons by the defects, 
implying the non-observation of such ultra-high-energy photons. Thus, such models 
could be in agreement with the current experimental indications
of GZK cutoffs for both protons and photons~\cite{Auger,Augergamma}.

We conclude, therefore, that the impressive constraints of~\cite{libsigl} may be
evaded in at least three different ways. (1) In the string formation, stretching and decay
framework (\ref{string1}, \ref{string2}, \ref{string3}) that we now prefer~\cite{emnnewuncert},
the kinematics and the 1PI scattering amplitude for $\gamma + \gamma_{\rm B} \rightarrow e^+ e^-$
are identical with those in conventional QED. (2) In the metric deformation framework
(\ref{Finsler}) the kinematics assumed in~\cite{libsigl} are inapplicable if the recoil is
isotropic: $\ll u_i \gg = 0$. (3) An extended formalism is required in models with a low
string scale $M_s < E$, including a novel absorption mechanism for the GZK cutoff for photons.
In such models, there would be compatibility~\cite{emnnewuncert2} with the hints
of time delays associated with photon refraction found by the MAGIC~\cite{MAGIC2} and
other experiments~\cite{HESSFermi}, in particular if there is a low density of 
D-particle defects per string length 
at red-shifts $z= 0.03$~\cite{Dvoid}.

\section*{Acknowledgements}

The work of N.E.M. and J.E. is partially supported by the European Union
through the Marie Curie Research and Training Network \emph{UniverseNet}
(MRTN-2006-035863), and that of D.V.N. by
DOE grant DE-FG02-95ER40917.


\begin{thebibliography}{99}
\bibitem{UHELIV} For recent reviews see:
T.~Jacobson, S.~Liberati and D.~Mattingly,
 Annals Phys.\  {\bf 321}, 150 (2006)
 [arXiv:astro-ph/0505267];
 D.~Mattingly,
 Living Rev.\ Rel.\  {\bf 8}, 5 (2005)
 [arXiv:gr-qc/0502097],
and references therein.


\bibitem{GZK} K.~Greisen,
 Phys.\ Rev.\ Lett.\  {\bf 16}, 748 (1966);
G.~T.~Zatsepin and V.~A.~Kuzmin,
 JETP Lett.\  {\bf 4}, 78 (1966)
 [Pisma Zh.\ Eksp.\ Teor.\ Fiz.\  {\bf 4}, 114 (1966)].


\bibitem{Kifune}  T.~Kifune,
 Astrophys.\ J.\  {\bf 518}, L21 (1999)
 [arXiv:astro-ph/9904164].


\bibitem{PM} R.~J.~Protheroe and H.~Meyer,
 Phys.\ Lett.\  B {\bf 493}, 1 (2000)
 [arXiv:astro-ph/0005349].


\bibitem{EMNGZK} J.~R.~Ellis, N.~E.~Mavromatos and D.~V.~Nanopoulos,
 Phys.\ Rev.\  D {\bf 63}, 124025 (2001)
 [arXiv:hep-th/0012216].


\bibitem{Dfoam2} J.~R.~Ellis, N.~E.~Mavromatos and D.~V.~Nanopoulos,
Gen.\ Rel.\ Grav.\  {\bf 32}, 127 (2000)
[arXiv:gr-qc/9904068];
Phys.\ Rev.\  D {\bf 61}, 027503 (1999)
[arXiv:gr-qc/9906029];
Phys.\ Rev.\  D {\bf 62}, 084019 (2000)
[arXiv:gr-qc/0006004].

\bibitem{libsigl} L.~Maccione, S.~Liberati and G.~Sigl,
arXiv:1003.5468 [astro-ph.HE].

\bibitem{Auger} J.~Abraham {\it et al.}  [Pierre Auger Collaboration],
 Astropart.\ Phys.\  {\bf 29}, 243 (2008)
 [arXiv:0712.1147 [astro-ph]].

\bibitem{Augergamma}  J.~Abraham {\it et al.}  [Pierre Auger Collaboration],
 Phys.\ Lett.\  B {\bf 685}, 239 (2010)
 [arXiv:1002.1975 [astro-ph.HE]].





\bibitem{Dfoam} J.~R.~Ellis, N.~E.~Mavromatos and M.~Westmuckett,
Phys.\ Rev.\ D \textbf{70}, 044036 (2004) [arXiv:gr-qc/0405066];
\emph{ibid.} {\bf 71}, 106006 (2005)~.



\bibitem{emnnewuncert} J.~R.~Ellis, N.~E.~Mavromatos and D.~V.~Nanopoulos,
Phys.\ Lett.\  B {\bf 665}, 412 (2008)
[arXiv:0804.3566 [hep-th]];

\bibitem{li} T.~Li, N.~E.~Mavromatos, D.~V.~Nanopoulos and D.~Xie,
Phys.\ Lett.\  B {\bf 679}, 407 (2009)
[arXiv:0903.1303 [hep-th]].


\bibitem{AEMNS} G.~Amelino-Camelia, J.~R.~Ellis, N.~E.~Mavromatos and D.~V.~Nanopoulos,
 Int.\ J.\ Mod.\ Phys.\  A {\bf 12}, 607 (1997)
 [arXiv:hep-th/9605211];
G.~Amelino-Camelia, J.~R.~Ellis, N.~E.~Mavromatos, D.~V.~Nanopoulos and S.~Sarkar,
 Nature {\bf 393}, 763 (1998)
 [arXiv:astro-ph/9712103];
 J.~R.~Ellis, K.~Farakos, N.~E.~Mavromatos, V.~A.~Mitsou and D.~V.~Nanopoulos,
 Astrophys.\ J.\  {\bf 535}, 139 (2000)
 [arXiv:astro-ph/9907340].

\bibitem{feynman} R.~P.~Feynman, R.~B.~Leighton and M.~Sands, \emph{The Feynman
Lectures on Physics} Vol. \textbf{2}, (Addison-Wesley, Reading Mass. 1977).

\bibitem{mavro2010} For a review, see N.~E.~Mavromatos,
arXiv:0906.2712 [hep-th], Foundations of Physics in press [doi: 10.1007/s10701-009-9372-z], and references therein.

\bibitem{sussk1} N.~Seiberg, L.~Susskind and N.~Toumbas,
JHEP {\bf 0006}, 021 (2000)
[arXiv:hep-th/0005040];

\bibitem{pioline} I.~Antoniadis and B.~Pioline,
 Nucl.\ Phys.\  B {\bf 550}, 41 (1999)
 [arXiv:hep-th/9902055].

\bibitem{burgess}  C.~P.~Burgess,
Nucl.\ Phys.\  B {\bf 294}, 427 (1987).

\bibitem{emnnewuncert2} J.~Ellis, N.~E.~Mavromatos and D.~V.~Nanopoulos,
Phys.\ Lett.\  B {\bf 674}, 83 (2009)
[arXiv:0901.4052 [astro-ph.HE]].

\bibitem{MAGIC2}  J.~Albert {\it et al.}  [MAGIC Collaboration] and
J.~R.~Ellis, N.~E.~Mavromatos, D.~V.~Nanopoulos, A.~S.~Sakharov and E.~K.~G.~Sarkisyan,
Phys.\ Lett.\  B {\bf 668}, 253 (2008).

\bibitem{HESSFermi}
F.~Aharonian {\it et al.},
  Phys.\ Rev.\ Lett.\  {\bf 101}, 170402 (2008)
  [arXiv:0810.3475 [astro-ph]];
  Fermi GBM and LAT Collaborations,
  arXiv:0908.1832 [astro-ph.HE].
  
\bibitem{Dvoid}
J.~Ellis, N.~E.~Mavromatos and D.~V.~Nanopoulos,
arXiv:0912.3428 [astro-ph.CO].


\end{thebibliography}
\end{document}